\documentclass[12pt,preprint]{aastex}

\usepackage{amsmath}
\usepackage{natbib}

\newcommand{\kms}{km s$^{-1}$}
\newcommand{\msun}{M$_{\odot}$}

\newcommand{\pmpc}{Mpc$^{-1}$}
\newcommand{\ntot}{541}
\newcommand{\ngal}{451}
\newcommand{\nhvc}{90}
\def\be{\begin{equation}}
\def\ee{\end{equation}}

\shorttitle{ALFALFA Catalog at $\delta=+25^{\circ}$}
\shortauthors{Martin et al.}

\begin{document}

\title{The Arecibo Legacy Fast ALFA Survey: VIII. HI Source Catalog of the Anti-Virgo Region at $\delta=+25^{\circ}$}

\author {Ann Martin\altaffilmark{1}, 
Riccardo Giovanelli\altaffilmark{1,2}, Martha P. Haynes\altaffilmark{1,2}, 
Am\'{e}lie Saintonge\altaffilmark{3}, G. Lyle Hoffman\altaffilmark{4}, Brian R. Kent\altaffilmark{5}, Sabrina Stierwalt\altaffilmark{1}}
\altaffiltext{1}{Center for Radiophysics and Space Research, Space Sciences Building,
Cornell University, Ithaca, NY 14853. {\textit{e-mail:}} amartin@astro.cornell.edu,
riccardo@astro.cornell.edu, haynes@astro.cornell.edu, sabrina@astro.cornell.edu}
\altaffiltext{2}{National Astronomy and Ionosphere Center, Cornell University,
Ithaca, NY 14853. The National Astronomy and Ionosphere Center is operated
by Cornell University under a cooperative agreement with the National Science
Foundation.}
\altaffiltext{3}{Insitute for Theoretical Physics, University of Zurich, Winterhurerstrasse 190, CH-8057 Zurich, Switzerland. {\textit{e-mail:}} amelie@physik.uzh.ch}
\altaffiltext{4}{Department of Physics, Lafayette College, Easton, PA 18042. {\textit{e-mail:}} hoffmang@lafayette.edu}
\altaffiltext{5}{Jansky Fellow of the National Radio Astronomy Observatory, 520 Edgemont Road, Charlottesville, VA 22903. NRAO is operated by Associated Universities, Inc., under cooperative agreement with the National Science Foundation. {\textit{e-mail:}} bkent@nrao.edu}

\begin{abstract}
We present a fourth catalog of HI sources from the Arecibo Legacy Fast ALFA (ALFALFA) Survey. We report \ntot \ detections over 136 deg$^2$, within the region of the sky having $22{\rm h}<\alpha<03{\rm h}$ and $24^{\circ}<\delta<26^{\circ}$. This complements a previous catalog in the region $26^{\circ}<\delta<28^{\circ}$ \citep[]{alfalfa5}. We present here the detections falling into three classes: (a) extragalactic sources with $S/N>6.5$, where the reliability of the catalog is better than 95$\%$; (b) extragalactic sources $5.0<S/N<6.5$ and a previously measured optical redshift that corroborates our detection; or (c) High Velocity Clouds (HVCs), or subcomponents of such clouds, in the periphery of the Milky Way. Of the \ntot \ objects presented here, \nhvc \ are associated with High Velocity Clouds, while the remaining \ngal \ are identified as extragalactic objects. Optical counterparts have been matched with all but one of the extragalactic objects. 
\end{abstract}

\section{Introduction}

First-generation blind HI surveys conducted in recent years (HI Parkes All Sky Survey, HIPASS: Barnes et al. 2001; Arecibo Dual-Beam Survey, ADBS: Rosenberg and Schneider 2000) have uncovered populations of HI-rich galaxies. Such surveys have made attempts at measuring the distribution of neutral hydrogen in the local Universe through such statistical measures as the HI mass function \citep{hipasshimf,adbshimf} and the galaxy-galaxy correlation function \citep{hipasscorr} of HI-selected objects. However, it is necessary for second-generation surveys to develop a larger sample that covers a cosmologically significant volume and a variety of environments in order to truly probe the statistical nature of this distribution without bias. Furthermore, the first-generation attempts suffer from small sample size, especially for low-mass objects, and low median sample redshifts. The Arecibo Legacy Fast ALFA Survey (ALFALFA), by contrast, with larger bandwidth, superior sensitivity, and high spatial resolution, is expected to make a significant contribution to the local HI census.

ALFALFA is an ongoing project at the Arecibo Observatory 305m telescope, taking advantage of the seven-beam ALFA receiver to conduct a large-scale blind survey of neutral hydrogen in the local Universe. The use of HI as a cosmological probe depends critically upon blind surveys, in order to prevent bias toward highly luminous objects. As a second-generation project, ALFALFA proposes to improve upon the results of HIPASS and ADBS and pave the way for large galaxy redshift surveys using the 21cm line. As the first survey of this kind to probe a cosmologically significant volume, approximately $2.7 \times 10^{7}$ Mpc$^{3}$, ALFALFA will provide a complete picture of the distribution of hydrogen in nearby galaxies. Additionally, the large sample size will allow studies of subsamples to investigate environmental influence; by design, ALFALFA will include galaxies from a range of environments, from the Pisces-Perseus foreground void \citep{alfalfa5}, to medium-density groups like Leo \citep{alfalfa8}, to the turbulent high-density Virgo Cluster \citep{alfalfa3,alfalfa6}.

ALFALFA will make neutral hydrogen measurements of $>25,000$ galaxies \citep[]{alfalfa1}. Tracing the distribution of HI out to $z \sim 0.06$, ALFALFA samples have a median redshift of about 7,000 \kms, compared to the median redshift of HIPASS at 3,000 \kms\ \citep{hipasscat}. Furthermore, ALFALFA probes deeper into the mass distribution, sensitive to neutral hydrogen masses of $M_{HI}\sim 2 \times 10^{6} M_{\odot}$ out to the distance of the Virgo Cluster, $\sim$16 Mpc. We expect to detect on the order of hundreds of objects with HI masses less than $10^{7.5} M_{\odot}$, a population of galaxies that has never before been adequately sampled. ALFALFA will thus provide a unique dataset for HI cosmology, including the measurement of the HI mass function (HIMF), especially at the low-mass end, and the correlation function of HI-selected galaxies at $z \sim 0$. Additionally, the ALFALFA footprint has significant overlap with galaxy samples from SDSS, 2MASS, and GALEX. Combining an HI survey with these datasets provides a more complete view of the relationship between star formation, environment, and gas. The small beam size and high sensitivity of the Arecibo Telescope, as compared with the Parkes Telescope used for HIPASS, reduces centroiding error, so that HI detections can be matched and cross-catalogs produced with great confidence in order to take full advantage of publicly available datasets at other wavelengths.

When complete, the ALFALFA survey will cover 7000 deg$^2$ of high galactic latitude sky out to $cz_{\odot}\sim 18,000$ \kms. This area is divided into two regions: a ``spring" sky as viewed from Arecibo, covering $07^{h}30^{m}<R.A.<16^{h}30^{m}$ and $0^{\circ}<\delta<+36^{\circ}$, and a ``fall" sky covering the same region in declination and $22^{h}00^{m}<R.A.<03^{h}00^{m}$. The spring sky includes the Virgo Cluster overdensity \citep{alfalfa3,alfalfa6}; here, we report on a 2$^{\circ}$-thick slice of the fall sky, described in \S2.

This work is the fourth catalog of ALFALFA sources published since observations commenced in February 2005. Previous catalogs \citep[]{alfalfa3,alfalfa5,alfalfa6}, and a fifth catalog paper currently submitted to the AJ \citep[]{alfalfa8}, have contained 706, 439, 564, and 546 extragalactic HI detections in regions covering 132, 135, 132 and 118 deg$^2$, respectively. Including the catalog presented here, ALFALFA has covered 653 deg$^2$, $\sim9\%$ of its total area and recovered 2706 HI-rich galaxies, or roughly $4.1$ sources deg$^{-2}$. For comparison, HIPASS detected 5317 extragalactic sources over 29,000 deg$^2$ \citep{hipasscat,northhipass} for an average of $\sim0.2$ sources deg$^{-2}$.

The remainder of this paper is organized as follows. In \S2, we provide an overview of the ALFALFA data collection methods, including our observing strategy and data reduction processes. A catalog of \ntot \ HI detections at $\delta=+25^{\circ}$ is presented in \S3. In \S4, we discuss several objects of note, followed by a discussion of the catalog's statistical properties and a summary of findings in \S5. Throughout, a value for the Hubble constant of 70 \kms \pmpc \ has been assumed.

\section{Data}

The catalog presented here covers 136 deg$^2$ of contiguous area in a 2$^{\circ}$-wide strip of the ALFALFA sky centered on $\delta=+25^{\circ}$. As part of ALFALFA's ``fall sky" described in \S1, the region stretches from $22^{h}00^{m}$ to $03^{h}00^{m}$ in right ascension. This region includes two key features: a portion of the Pisces-Perseus Supercluster (PPS), including the southernmost region of the prominent main ridge of the cluster, and a portion of a void in the foreground of the PPS (first reported in \citet{hg86}) around $cz\lesssim2500$ \kms.

ALFALFA observations are conducted at the Arecibo telescope, using the 7-pixel ALFA receiver installed in 2004. The observing strategy is minimally invasive, employing a fixed-azimuth drift scan mode. As a result of this strategy, many nights of observing time are required to fill in the full declination range. Additionally, because two passes are made over every point in the survey, a given region of sky will typically not be completed until many months after it is first observed. Since this second pass occurs at a different point in the Earth's orbit around the Sun, and Doppler tracking is not employed, this strategy more clearly reveals spurious signals resulting from radio frequency interference (RFI) while confirming true cosmic sources. ALFA's beams are elliptical in shape, resulting in an angular resolution of the survey of $3.3^{\prime} \times 3.8^{\prime}$. For a galaxy at a distance of 10 Mpc, this corresponds to a physical scale of 10 kpc; for a galaxy at the very edge of the ALFALFA volume, at 250 Mpc, this corresponds to a physical scale of 250 kpc. We cover a bandwidth of 100 MHz (-1,600 \kms\ to 18,000 \kms) with a spectral resolution, before spectral smoothing, of 5.5 \kms. 

Once the data is acquired for a given region, individual 600-second drift scans are combined into three-dimensional data cubes known as ``grids." These cubes cover 2.4$^{\circ}$ in both R.A. and declination, with 1\arcmin \ sampling, and are designed to overlap so that sources on the edges of grids are easily recoverable. The sources for this catalog were obtained from 38 such grids. Each grid is also broken down into four ``subgrids" along the spectral direction, again with substantial overlap. Each of these subgrids covers 1024 channels along the spectral axis, respectively covering the ranges -2,000 to 3,300 \kms, 2,500 to 7,900 \kms, 7,200 to 12,800 \kms, and 12,100 to 17,900 \kms. The grids also provide useful information on the positions of continuum sources. By comparing the positions measured by ALFALFA to those given in the NVSS catalog \citep{nvss}, we are able to account for Arecibo's systematic pointing errors. This process results in excellent position reliability for the sources extracted from each grid. The processing undergone by each grid is discussed in detail in \citet{alfalfa3} and \citet{alfalfa4}.

Source detection and extraction uses a combination of automated techniques, which reliably identify a set of possible signals, and individual assessment and measurement of confirmed sources by the authors. Candidate detections are identified using a matched-filtering protocol with galaxy profile templates built from combinations of Hermite functions \citep[]{alfalfa4}. Each of the resulting candidates is examined individually, and for those selected for inclusion in the final catalog, a set of source parameters is interactively measured. These parameters are presented in \S3. Each source is also compared with optical images from DSS2 and cross-correlated with NED and the AGC (``Arecibo General Catalog," a private database of galaxies maintained by M.P.H. and R.G.). Some Sloan Digital Sky Survey coverage is now available for this region, following the seventh data release of SDSS imaging and spectroscopy (DR7). The SDSS data was used to confirm DSS2 optical information in a few ambiguous cases. While most galaxies in this catalog have a high signal-to-noise ratio and can be confidently included on the basis of their HI profile alone, 21 of these sources are labeled as "Code 2" objects (see \S3), which have slightly lower $S/N$ (typically $5.0<S/N<6.5$) but are well-corroborated by previous redshift measurements. Each galaxy is confidently matched with an optical counterpart with only one exception (see \S4). These matches are mostly unambiguous, and the positions of the HI centroid and optical center match very well, although the pointing offsets are dependent on $S/N$. For all detections in the catalog, the median pointing offset between the measured HI coordinates and the assigned optical counterpart, is 18\arcsec \, with the error reduced to 14\arcsec \ for detections with  $S/N>12$. The absolute HI positions themselves are calculated using a fit from continuum maps \citep{alfalfa6}; continuum sources detected in ALFALFA are compared to VLA radio continuum sources (NVSS; \citet{nvss}) to correct Arecibo pointings, thus removing this source of systematic error from ALFALFA coordinates. The typical RMS for all sources is $\sim2.3$ mJy.	

Our point source mass sensitivity as a function of distance, and related scaling relations, are detailed in \citet{alfalfa1}; for a particular signal--to--noise ratio S/N our mass sensitivity is:

        \begin{equation}
        \frac{M_{HI}}{M_{\odot}}=2.356 \times 10^5 \: D_{Mpc}^2 \: F_c =2.356\times 10^5 \: D_{Mpc}^2 \: \frac{W50}{w_{smo}^{1/2}} \: \frac{(S/N) \: \sigma_{rms}}{1000}
	\label{eqsn}
	\end{equation}

where $F_c$ is the integrated flux density in Jy \kms, $W50$ is the velocity width of the line profile at the 50\% level, $w_{smo}$ is a smoothing width (either $W50/20$ for $W50<$ 400 \kms \ or 20 for $W50 \ge$ 400 \kms), and $\sigma_{rms}$ is the r.m.s. noise figure measured in mJy at 10 \kms \ resolution. These parameters are more thoroughly discussed in \S3, as is the relationship between integrated flux and mass sensitivity.

\section{Catalog Presentation \label{cat}}

\begin{deluxetable}{cccccccccccccc}
\rotate
\tablewidth{0pt}
\tabletypesize{\tiny}
\tablecaption{HI Candidate Detections\label{params}}
\tablehead{
\colhead{Source}  & \colhead{AGC}   & \colhead{$\alpha_{J2000}$ (HI)} & \colhead{$\delta_{J2000}$ (HI)} &
\colhead{$\alpha_{J2000}$ (Opt.)} & \colhead{$\delta_{J2000}$ (Opt.)} &
\colhead{cz$_\odot$}  & \colhead{$w50 ~(\epsilon_w$)} &
\colhead{$F_{c}$} & \colhead{S/N} & \colhead{rms} &
\colhead{Dist}    & \colhead{$\log M_{HI}$} & \colhead{Code}
    \\
 & & & & & &{\kms} & {\kms} & {Jy \kms} & & {mJy} & Mpc & {$M_\odot$} &
    \\
 Column 1&2&3&4&5&6&7&8&9&10&11&12&13&14
}
\startdata
4-  1  & 321318 & 22 00 55.4 & +24 45 10 &            &           &   -345 &    29( 14) &   2.00 &   11.4 &   7.01 &        &        & 9 * \\
4-  2  & 321319 & 22 01 27.2 & +24 59 47 &            &           &   -371 &    20(  9) &   1.08 &    9.8 &   5.07 &        &        & 9 * \\
4-  3  & 321303 & 22 02 06.7 & +24 15 32 & 22 02 05.3 & +24 15 32 &   6059 &   127( 18) &   1.66 &   15.4 &   2.13 &   88.4 &   9.49 & 1   \\
4-  4  & 321304 & 22 03 51.1 & +25 26 59 & 22 03 51.1 & +25 26 32 &   2692 &   142( 18) &   1.71 &   15.7 &   2.04 &   41.3 &   8.84 & 1   \\
4-  5  & 320941 & 22 07 50.7 & +24 50 01 & 22 07 51.0 & +24 49 57 &   3412 &   124(  9) &   3.71 &   38.6 &   1.92 &   51.4 &   9.36 & 1   \\
4-  6  & 320047 & 22 08 50.1 & +24 42 09 & 22 08 49.2 & +24 41 46 &   6300 &   284( 40) &   1.14 &    8.0 &   1.90 &   91.6 &   9.35 & 1   \\
4-  7  & 320063 & 22 10 15.2 & +25 28 11 & 22 10 16.7 & +25 27 57 &   7144 &   124(  5) &   1.25 &   11.1 &   2.26 &   97.2 &   9.44 & 1   \\
4-  8  & 321305 & 22 10 21.6 & +24 44 11 & 22 10 22.4 & +24 43 39 &  12185 &   137( 26) &   0.85 &    8.0 &   2.03 &  169.2 &   9.76 & 1   \\
4-  9  & 321306 & 22 11 37.8 & +25 43 35 & 22 11 37.7 & +25 43 46 &  11875 &   135( 16) &   0.83 &    7.2 &   2.21 &  164.8 &   9.73 & 1   \\
4- 10  & 321320 & 22 12 27.6 & +24 43 30 &            &           &   -303 &    25(  8) &   1.83 &   14.2 &   5.46 &        &        & 9 * \\
4- 11  & 321204 & 22 13 05.5 & +25 54 37 & 22 13 06.2 & +25 54 56 &  12652 &   133( 24) &   1.10 &    9.4 &   2.27 &  175.9 &   9.90 & 1   \\
4- 12  & 321307 & 22 14 04.4 & +25 41 08 & 22 14 04.7 & +25 40 52 &   1152 &    60(  5) &   1.04 &   15.6 &   1.89 &   18.7 &   7.94 & 1   \\
4- 13  & 321309 & 22 16 09.2 & +25 18 10 & 22 16 08.7 & +25 18 16 &   5189 &   157( 18) &   0.81 &    8.1 &   1.78 &   76.1 &   9.04 & 1   \\
4- 14  & 321368 & 22 16 56.9 & +25 33 49 &            &           &   -459 &    18(  6) &   0.43 &    6.8 &   3.09 &        &        & 9 * \\
4- 15  & 320128 & 22 17 13.6 & +25 12 32 & 22 17 13.1 & +25 12 47 &  12654 &   412( 15) &   2.46 &   14.0 &   1.90 &  175.9 &  10.25 & 1   \\
4- 16  & 321321 & 22 17 20.3 & +25 42 31 & 22 17 20.2 & +25 42 27 &  12078 &    39(  7) &   0.94 &   15.0 &   2.19 &  167.7 &   9.79 & 1   \\
4- 17  & 320139 & 22 18 05.3 & +25 05 53 & 22 18 07.2 & +25 05 32 &   7049 &    78( 12) &   0.70 &    8.8 &   1.97 &   95.8 &   9.18 & 1   \\
4- 18  & 321369 & 22 18 14.6 & +24 42 29 &            &           &   -424 &    31(  8) &   1.32 &   16.8 &   3.06 &        &        & 9 * \\
4- 19  & 320149 & 22 19 06.5 & +24 35 37 & 22 19 06.2 & +24 35 53 &   8247 &   178( 23) &   0.97 &    8.7 &   1.86 &  112.9 &   9.46 & 1   \\
4- 20  & 321350 & 22 19 38.4 & +25 47 36 &            &           &   -427 &    18(  2) &   1.60 &   16.1 &   4.85 &        &        & 9 * \\
4- 21  & 321322 & 22 19 46.9 & +25 41 52 & 22 19 47.2 & +25 41 32 &   8110 &   145( 24) &   0.98 &    9.8 &   1.84 &  111.0 &   9.45 & 1   \\
4- 22  & 321323 & 22 20 06.3 & +25 36 31 & 22 20 06.6 & +25 36 20 &  12482 &   357( 14) &   2.82 &   14.8 &   2.26 &  173.4 &  10.30 & 1   \\
4- 23  & 321389 & 22 20 23.2 & +25 10 33 &            &           &   -428 &    31(  5) &   0.77 &    8.3 &   3.58 &        &        & 9 * \\
4- 24  & 321324 & 22 20 22.9 & +24 23 14 & 22 20 20.6 & +24 23 18 &   3971 &   147( 22) &   2.02 &   18.2 &   2.04 &   58.9 &   9.22 & 1   \\
4- 25  & 321325 & 22 20 26.8 & +25 47 06 & 22 20 23.9 & +25 47 16 &  11962 &   396( 10) &   1.12 &    6.5 &   1.95 &  166.0 &   9.86 & 2   \\
4- 26  & 320162 & 22 20 44.9 & +24 46 48 & 22 20 41.3 & +24 45 52 &  12349 &   239( 13) &   1.01 &    7.7 &   1.89 &  171.5 &   9.85 & 1   \\
4- 27  & 321326 & 22 20 54.3 & +24 45 31 & 22 20 54.0 & +24 45 50 &   5256 &   166( 35) &   1.35 &   11.0 &   2.14 &   76.9 &   9.27 & 1   \\
4- 28  & 321351 & 22 21 30.5 & +24 36 58 &            &           &   -341 &    26(  5) &   0.64 &    8.4 &   3.22 &        &        & 9 * \\
4- 29  & 321352 & 22 21 45.3 & +25 49 29 &            &           &   -440 &    21(  2) &   5.79 &   48.7 &   5.47 &        &        & 9 * \\
4- 30  & 321390 & 22 22 13.9 & +24 33 53 &            &           &   -420 &    24(  4) &   0.99 &    9.6 &   4.43 &        &        & 9 * \\
4- 31  & 321328 & 22 23 13.8 & +24 38 01 & 22 23 12.5 & +24 38 21 &  12314 &   307( 38) &   2.26 &   14.4 &   2.00 &  171.0 &  10.19 & 1   \\
4- 32  & 321391 & 22 24 17.1 & +25 33 50 &            &           &   -403 &    29(  5) &   0.60 &    8.2 &   2.93 &        &        & 9 * \\
4- 33  & 321327 & 22 24 24.0 & +24 15 06 & 22 24 25.2 & +24 14 27 &  12072 &   229( 19) &   1.36 &   10.2 &   1.96 &  167.5 &   9.95 & 1   \\
4- 34  & 321225 & 22 24 53.4 & +25 57 31 &            &           &   -407 &    22(  2) &   0.92 &   12.5 &   3.33 &        &        & 9 * \\
4- 35  & 321329 & 22 26 26.4 & +25 22 32 & 22 26 25.0 & +25 22 27 &  11954 &    81(  9) &   0.81 &   10.1 &   1.97 &  165.9 &   9.72 & 1   \\
4- 36  & 321330 & 22 26 51.9 & +25 00 03 & 22 26 50.3 & +25 00 57 &   7308 &    95(  8) &   0.72 &    6.7 &   2.44 &   99.5 &   9.22 & 1   \\
4- 37  & 321331 & 22 27 30.5 & +25 37 50 & 22 27 31.5 & +25 37 40 &  12191 &   226( 26) &   1.09 &    8.1 &   1.98 &  169.2 &   9.87 & 1   \\
4- 38  & 321332 & 22 28 14.1 & +25 46 54 & 22 28 13.2 & +25 47 11 &   7254 &    30(  6) &   0.56 &   10.7 &   2.04 &   98.7 &   9.11 & 1   \\
4- 39  & 321394 & 22 29 17.7 & +25 00 53 &            &           &   -289 &    36(  9) &   0.37 &    6.5 &   2.09 &        &        & 9 * \\
4- 40  & 321333 & 22 30 00.5 & +25 09 02 & 22 29 59.2 & +25 09 00 &  11994 &   224( 16) &   1.11 &    8.4 &   1.97 &  166.4 &   9.86 & 1   \\
\hline
\enddata
\end{deluxetable}

We present in Table \ref{params} the measured parameters for \ntot \ detctions, \ngal \ of which are associated with extragalactic objects, while the remaining \nhvc \ are HVC features. The contents of the different columns are:
\begin{itemize}

\item Col. 1: an entry number for this catalog

\item Col. 2: the source number in the Arecibo General Catalog, a private database
        of extragalactic objects maintained by M.P.H. and R.G. AGC numbers, along with all other parameters, will be made available on our public digital 
	archive site\footnotemark \footnotetext{http://arecibo.tc.cornell.edu/hiarchive/alfalfa/} and are listed on the NASA Extragalactic Database\footnotemark 
	\footnotetext{http://nedwww.ipac.caltech.edu/}(NED). Therefore, the AGC, while a private databasse, provides a single, unique identifier for these HI
	sources and their optical counterparts.


\item Cols. 3 \& 4: center (J2000) of the HI source, after correction for systematic 
	telescope pointing errors, which are on the order of 20\arcsec \ and depend on declination.
	The accuracy of the HI positions, compared to the positions of the optical counterparts, depends on source strength. On average,
	the positional accuracy, estimated as the difference between the HI position and the optical counterpart, is about 18\arcsec. 

\item Cols. 5 \& 6: center (J2000) of the optical counterpart matched with the source. No optical counterpart is listed for High Velocity Clouds.
	The position has been checked for each listed object and assessed using tools provided
	through the {\it SkyView} website, in addition to NED and the AGC. The quality of centroids
	is estimated to be 2\arcsec~ or better. One extragalactic object has no identified optical countpart.
	In the case of several other objects with a listed optical counterpart, a note explains an ambiguity, as alerted
	by an asterisk in Col. 14.

\item Col. 7: heliocentric velocity of the HI source, $cz_{\odot}$, measured 
	as the midpoint 
	between the channels at which the flux density drops to 50\% of 
	each of the two peaks (or of one, if only one is present) at each
	side of the spectral feature. Units are \kms. The error on $cz_\odot$
	to be adopted is half the error on the width, tabulated in Col. 8.

\item Col. 8: velocity width of the source line profile, $W50$, measured at the 50\%
	level of each of the two peaks, as described for Col. 7. This value 
	is corrected for instrumental broadening. No corrections due to
	turbulent motions, disk inclination or cosmological effects are
	applied. In parentheses we show the estimated error on the velocity width, $\epsilon_w$, in \kms.
	This error is the sum in quadrature of two components: the first is a
	statistical error; the second is a systematic error
	associated with the subjective guess with which the observer estimates 
	the spectral boundaries of the feature, flagged during the interactive assessment of candidate detections. In the majority of cases,
	the systematic error is significantly smaller than the statistical
	error; thus the former is ignored.

\item Col. 9: integrated flux density of the source, $F_c$, in Jy \kms . This 
	is measured on the integrated spectrum, obtained by
	spatially integrating the source image over a solid angle of at
        least $7$\arcmin $\times 7$\arcmin ~and dividing by the sum of the survey beam
	values over the same set of image pixels \citep[see][]{shostak}. 
	Integrated fluxes for very extended sources with large
	spatial asymmetries can be misestimated by our 
	algorithm, which is not optimized for that category of sources. A special catalog with parameters
	of extended sources will be produced after completion of the survey. 

	The issue is especially severe for extended High Velocity Clouds 
	that exceed in size that of the ALFALFA data cubes. In these specific 
	cases, concentrations of emission are identified and the flux in these knots of emission are measured in the same way as 
	extragalactic sources and included as separate entries in Table 1.  
	In general, this meant 	applying the same 
	kind of $S/N$ selection threshold as for the extragalactic signals, with 
	the exception of the southern extension of Wright's cloud \citep{wright}. This cloud extends into the region
	covered by \citet{alfalfa5} and was discussed in that previous ALFALFA data release. The cloud has a significant
	velocity gradient over its full extent, from about -350 \kms / to -475 \kms, and is thought to
	potentially be associated with M33.
	Here, we have made a bulk measurement of the region of the cloud
	extending South of +26$^{\circ}$ in declination, in addition to separate measurements of a selection of the brightest knots. 
        
	Generally, our measurements of HVCs are likely to be underestimates of the total size and flux, since we are
	not very sensitive to diffuse emission that may connect filaments and fill the most extended clouds. Detailed study
        of extended HI features and HVCs will therefore take place in a future paper.
	See Column 14 
	and the corresponding comments for individual objects. 

\item Col. 10: signal--to--noise ratio S/N of the detection, estimated as 
        \begin{equation}
	S/N=\left ( \frac{1000F_c}{W50} \right ) \frac{w_{smo}^{1/2}}{\sigma_{rms}}
	\label{eqsn}
	\end{equation}
        where $F_c$ is the integrated flux density in Jy \kms, as listed in Col. 9;
        the ratio $1000 F_c/W50$ is the mean flux across the feature in mJy;
        $w_{smo}$ is either $W50/(2\times 10)$ for $W50<400$ \kms \ or
        $400/(2\times 10)=20$ for $W50 \geq 400$ \kms ($w_{smo}$ is a
        smoothing width expressed as the number of spectral resolution
        bins of 10 \kms \ bridging half of the signal width); and $\sigma_{rms}$
        is the r.m.s noise figure across the spectrum measured in mJy at 10
	\kms \ resolution, as tabulated in Col. 11.

\item Col. 11: noise figure of the spatially integrated spectral profile, $\sigma_{rms}$,
	in mJy. The noise figure as tabulated is the r.m.s. as measured over the signal-- and
	rfi-free portions of the spectrum, after Hanning smoothing to a spectral
	resolution of 10 \kms. The regions of the spectrum affected by rfi are identified, based on visual inspection by a member of the ALFALFA collaboration,
	in an early stage of data processing. These identifications are then tracked through the data processing pipeline, allowing
	affected spectral channels to be excluded from the calculation of the noise figure, as discussed in \citet{alfalfa3}. Prominent sources of rfi include the Federal
	Aviation Administration radar operating at 1350 MHz from the San Juan Airport.

\item Col. 12: adopted distance in Mpc, $D_{Mpc}$. For objects with $cz_{\odot}>6000$ \kms, 
	the distance is simply estimated as $cz_{cmb}/H_\circ$; $cz_{cmb}$ is the recession velocity
	measured in the Cosmic Microwave Background reference frame \citep{vhel2cmb} and $H_\circ$ is
	the Hubble constant, for which we use a value of 70 \kms Mpc$^{-1}$. For objects of lower
	$cz_{cmb}$, we use the local Universe peculiar velocity model of \citet{mastersth}, which is based on data
	from the SFI++ catalog of galaxies \citep{sfi++}. The \citet{mastersth} peculiar velocity model results
	from analysis of the peculiar motions of galaxies, groups, and clusters, using a combination of primary
	distances from the literature and secondary distances from the Tully-Fisher relation. The resulting model
	includes two attractors, with infall onto the Virgo Cluster and the Hydra-Centaurus Supercluster, as well as
	a quadrupole and a dipole component. The transition from one distance estimation method to the other
	is selected to be at $cz_{\odot}=6000$ \kms \ because the uncertainties in each method become comparable at that point.
	In cases where a galaxy has a known primary distance, that distance will be adopted; when the galaxy
	is a known member of a group \citep{sfi++}, that group's recessional velocity $cz_{cmb}$ is used to determine the distance
	estimate according to the prescription just described.
	
\item Col. 13: logarithm in base 10 of the HI mass, in solar units. That parameter is 
	obtained by using the expression $M_{HI}=2.356\times 10^5 D_{Mpc}^2 F_c$. 

\item Col. 14: object code, defined as follows: 
	
	Code 1 refers to sources 
	of high S/N and general qualities that make it a reliable detection. These signals exhibit
	a good match between the two independent polarizations observed by ALFALFA,
	a spatial extent consistent with the telescope beam, a spectral profile clean of RFI features,
	and an approximate S/N threshold of 6.5. These
	criteria lead to the exclusion of some candidate detections with $S/N>6.5$; likewise, some
	features with S/N slightly below this soft threshold are included, due to optimal overall characteristics
	of the feature, such as well-defined spatial extent, broad velocity width, and obvious association with an optical
	counterpart. We estimate that the detections with code 1 in Table \ref{params} will be confirmed with follow-up
	observations in greater than 95$\%$ of cases \citep{alfalfa4}.

	Code 2 refers to the sources we call `priors.' These are sources of low S/N ($<$ 6.5), which would  
        ordinarily not be considered
	reliable detections by the criteria set for code 1, but have been matched with optical counterparts with
	known optical redshifts which corroborate that measured in the HI line. We include them in our catalog because
	they are very likely to be real. There are 21 such sources in the present catalog.

	Code 9 refers to objects assumed to be HVCs; no
	estimate of their distances is made.

 	Notes flag. An asterisk in this column indicates a comment is included
	for this source in the text below.

\end{itemize}

Only the first few entries of Table 1 are listed in the printed version of this
paper. The full content of Table 1 is accessible through the electronic version
of the paper and will be made available also through our public digital 
archive site\footnotemark \footnotetext{http://arecibo.tc.cornell.edu/hiarchive/alfalfa/}. 
The comments for those sources marked with an asterisk in column 14 are given here:\\
\\
\footnotesize
 4-  1: HVC: small cloud \\
 4-  2: HVC: small cloud \\
 4- 10: HVC: small, isolated cloud \\
 4- 14: HVC: smaller, relatively isolated cloud in a densely populated region. \\
 4- 18: HVC: bright knot of a larger cloud in a densely populated region. \\
 4- 20: HVC: bright knot in an elongated filament in a densely populated region. \\
 4- 23: HVC: bright knot in filament in densely populated region. \\
 4- 28: HVC: more isolated cloud in velocity, but near a densely populated region. \\
 4- 29: HVC: bright peak in a large cloud in a densely populated region. \\
 4- 30: HVC: bright knot in HVC field/filament. \\
 4- 32: HVC: small knot in relatively dense field. \\
 4- 34: HVC: elongated cloud in somewhat dense field. \\
 4- 39: HVC: small, isolated cloud \\
 4- 42: HVC: large, elongated cloud \\
 4- 44: HVC: cloud in dense region \\
 4- 45: cz mismatch with previous detection (Lawrence et al. 1999, MNRAS, 308, 897), but within quoted measurement error \\
 4- 47: HVC: small, more isolated cloud near somewhat dense region \\
 4- 52: poor width precision due to shape of HI profile and poor spectral definition \\
 4- 53: signal merges into strong rfi, severely affecting parameters \\
 4- 55: blend of emission from AGC320702 (224005.8+244156) and AGC320701 (223957.0+244139); not separable spatially or kinematically. Parameters uncertain. \\
 4- 58: HVC: small cloud \\
 4- 59: HVC: small, relatively isolated cloud not far from some others \\
 4- 60: HVC: small, relatively isolated cloud in a field with several others \\
 4- 62: HVC: clumpy/patchy cloud in a region near others \\
 4- 70: HVC: bright knot in elongated filament in densely populated region \\
 4- 71: HVC: bright knot in elongated filament in densely populated region \\
 4- 73: HVC: bright peak in elongated cloud, in densely populated region \\
 4- 75: HVC: bright peak in cloud complex \\
 4- 76: HVC: bright peak in cloud complex \\
 4- 77: HVC: bright knot in cloud complex \\
 4- 78: HVC: bright knot in cloud complex \\
 4- 80: HVC: bright knot in elongated filament, part of a larger cloud complex \\
 4- 86: blend with emission from UGC12290; deblending good, but centroiding and parameters for AGC320603 moderately uncertain. \\
 4- 91: alternative opt.id with 230003.9+253359 (extremely LSB); position poorly determined. Associated with group SRGb016 (Mahdavi et al. 1999). \\
 4- 93: alternative opt.id with 230038.4+244704 \\
 4- 99: blend of emission from two nearby galaxies; deblending not possible. Lower-velocity emission is likely from the small nearby galaxy at 230212.7+244549; most of the emission is from the larger source. \\
 4-101: HVC: small cloud \\
 4-103: HVC: small cloud \\
 4-106: double system with partial deblending: detection includes some flux from neighbor, UGC12386. Parameters and centroiding affected. \\
 4-107: double system with partial deblending: detection includes some flux from neighbor, UGC12384. Parameters and centroiding affected. \\
 4-111: HVC: large cloud \\
 4-118: blend with emission from nearby AGC330159 (231529.1+250714); not separable spatially or kinematically. Parameters uncertain. \\
 4-123: blend with emission from AGC330217 (231840.2+251601) to the north and signs of interaction. Parameters and centroiding affected. \\
 4-125: HVC: elongated cloud \\
 4-128: HVC: small clump; somewhat difficult to distinguish from bulk of Galactic HI emission. \\
 4-136: note narrow profile width for object at cz = 9700 km/s \\
 4-141: HVC: small isolated cloud \\
 4-148: blend with emission from AGC331873 (232601.0+254137); signal merges into strong rfi, affecting parameters on high cz end of profile; deblending poor due to spatial and kinematic overlap \\
 4-150: HVC: small clump; somewhat difficult to distinguish from bulk of Galactic HI emission \\
 4-179: near region of poor spatial and spectral coverage; affects centroiding and distance between HI center and opt.id \\
 4-183: poor spectral definition on low-cz end of profile \\
 4-188: opt.id's appear to be an interacting system \\
 4-193: poor spectral definition, with ambiguous opt.id: possible alternative id with 234528.5+241303 and 234526.8+241307; emission may be a blend from all three \\
 4-195: HVC: large cloud; somewhat difficult to distinguish from bulk of Galactic HI emission \\
 4-204: HVC: small cloud \\
 4-207: possible alternative opt.id with 235449.2+245058, but less likely \\
 4-208: blend of AGC331012 with AGC331014 (235548.1+253031) and 331015 (235551.8+252933); no deblending. Parameters uncertain: signal is intrinsically broad, but also merges into rfi and approaches end of ALFALFA bandwidth. \\
 4-209: HVC: isolated large cloud \\
 4-216: HVC: small, isolated knot \\
 4-220: UGC94 nearby at same cz; deblending good, parameters mostly unaffected \\
 4-221: UGC89 nearby at same cz; deblending good, parameters mostly unaffected \\
 4-222: HVC: small, isolated cloud \\
 4-225: possible alternative opt.id with 001034.8+243005, farther from HI center. \\
 4-227: poor spectral definition (high noise) \\
 4-228: significant polarization mismatch \\
 4-234: possible alternative opt.id with 001611.8+245211 \\
 4-237: parameters affected by proximity to AGC100146 (001759.7+243345) \\
 4-238: parameters affected by proximity to UGC165 \\
 4-241: HVC: bright knot of a larger cloud \\
 4-248: opt.id ambiguous; other possible counterpart at 002258.5+254723 but farther from HI center \\
 4-249: near border of region; detection and parameters will be improved once data is available for +23 degree strip to the South \\
 4-255: HVC: bright knot of a larger cloud extending to the east \\
 4-257: HVC: bright knot \\
 4-260: HVC: bright knot of a much larger complex of clouds \\
 4-261: unresolved broad blend of emission from HCG 001. Centered on UGC248 pair (disturbed optical morphology); blend of flux from group, with uncertain parameters. Not separable spatially or kinematically. \\
 4-268: HVC: bright knot in an extended filament/cloud structure \\
 4-269: HVC: bright knot in a much larger complex of clouds \\
 4-272: HVC: bright knot in a much larger complex of clouds \\
 4-277: HVC: bright knot in a much larger complex of clouds \\
 4-280: possible alternative opt.id with 003237.2+244854, but much less likely \\
 4-281: HVC: one of the many knots in this densely populated region \\
 4-285: HVC: one of the many knots in this densely populated region. \\
 4-286: HVC: one of the many knots in this densely populated region. \\
 4-287: opt.id is distant (~1.6 arcmin), but has previous optical redshift measurement of cz=9838 km/s \\
 4-290: HVC: one of the many knots in this densely populated region. \\
 4-292: signal merges with strong rfi; parameters uncertain \\
 4-294: HVC: one of the many knots in this densely populated region. \\
 4-298: opt.id ambiguous; other possible counterpart at 003828.2+253616 \\
 4-308: parameters affected by poor coverage in this region \\
 4-313: possible alternative opt.id with 004433.2+254143 \\
 4-317: blend with AGC100511 (004557.3+251308); significantly broadens the profile \\
 4-322: alternative opt.id with 004842.0+243554; flux may be a blend. \\
 4-324: HVC: one of two measured peaks in a large cloud in this densely populated region \\
 4-327: blend of emission from AGC100594 and AGC100596 (005025.6+243114); not separable spatially or kinematically \\
 4-330: HVC: bright peak near a filament that blends into Galactic emission \\
 4-336: blend of emission from AGC101857 and AGC101858 (005456.4+255308); not separable spatially or kinematically \\
 4-337: HVC: bright peak in a densely populated region \\
 4-339: HVC: isolated, small cloud in a densely populated region \\
 4-342: alternative opt.id with 005629.1+241911 \\
 4-343: HVC: larger cloud in a densely populated region. \\
 4-344: HVC: compact cloud in a densely populated region \\
 4-346: HVC: bright peak in a large cloud, distinct from Galactic emission. \\
 4-352: HVC: large cloud of emission, not entirely spectrally distinct from bulk of Galactic HI emission. \\
 4-354: alternative opt.id with 010555.1+243435; emission may be a blend of both \\
 4-355: poor centroiding; galaxy emission plus a low-level emission extended in redshift space. Identified as AGC110025, but some emission is likely to be associated with nearby galaxies at 010619.0+253212.3 and 010617.4+253342.9. Unmeasured broader emission covers 6330 km/s to 6600 km/s, and the measurementof AGC110025 likely contains some contamination from this additional gas. \\
 4-358: HVC: subclump of Wright's Cloud complex \\
 4-359: HVC: subclump of Wright's Cloud complex \\
 4-360: HVC: entirety of the Wright's Cloud clump in +25deg region. Contains many sub-clumps which were mesasured separately; bulk of the cloud extends into +27deg region (see Saintonge et al. 2008) \\
 4-361: HVC: small clump, part of Wright's Cloud complex \\
 4-362: HVC: possibly an unusual positive velocity cloud. Significant polarization squint; caution as signal may be spurious. \\
 4-364: signal merges with strong rfi; parameters affected and some flux lost \\
 4-365: HVC: clump in an area with several more clouds \\
 4-368: HVC: clump in an area with several more clouds \\
 4-372: interacting triplet: UGC884, AGC113819 (012059.7+253207) and AGC110845 (012052.9+253247). Likely to be at a lower velocity; rfi strongly affects profile and parameters. Actual velocity width of emission closer to 400 km/s. \\
 4-377: HVC: large isolated cloud \\
 4-378: HVC: one of two clumps in a small faint cloud \\
 4-379: HVC: one of two clumps in a small faint cloud \\
 4-380: signal merges with strong rfi; parameters uncertain \\
 4-381: HVC: the brightest of several small distinct clouds in this region \\
 4-384: HVC: one of the small clouds in this region \\
 4-386: HVC: one of the small clouds in this region \\
 4-388: opt.id ambiguous: second possible counterpart at 013744.3+243153 \\
 4-389: HVC: cloud not completely spectrally distinct from bulk of Galactic HI emission \\
 4-390: HVC: companion to cloud HI013806.2+245528, not completely spectrally distinct from bulk of Galactic HI emission \\
 4-395: possible alternative opt.id with AGC113308 (014429.3+255114, no previous cz), but associated with AGC113307 because of match with previous measurement \\
 4-397: opt.id ambiguous; other possible counterpart at 014509.6+245535 \\
 4-401: opt.id ambiguous; other possible counterpart at 014619.6+252140 \\
 4-412: HVC: cloud not completely spectrally distinct from bulk of Galactic HI emission \\
 4-417: flux much larger than previous measurements (available through our public digital archive site) \\
 4-432: HVC: clumpy, irregular shape, well separated spectrally from galactic HI. Extends about half a degree; may extend into unavailable region to the south at +23deg \\
 4-440: HVC: central and brightest region of a filament that extends for over 2 degrees \\
 4-441: alternative opt.id with 021303.9+252126 \\
 4-446: HVC: group of knots \\
 4-447: parameters uncertain; signal may be contaminated by AGC121503 (021610.7+251413). Deviation from 2-horn profile shape and possible signs of interaction \\
 4-448: parameters uncertain; signal may be contaminated by UGC1739. Deviation from 2-horn profile shape and possible signs of interaction \\
 4-449: no discernible optical counterpart \\
 4-450: HVC: one of the many knots in this densely populated region \\
 4-451: HVC: patchy cloud that may extend farther south into unavailable region at +23deg \\
 4-452: galaxy shows clear signs of interaction: optical morphology very irregular, and presence of an extension of the HI, wider than the main emission from the galaxy \\
 4-453: HVC: one of the many knots in this densely populated region \\
 4-456: HVC: one of the many knots in this densely populated region \\
 4-458: HI disk seems elongated toward the NW, low surface brightness extension to one side of the galaxy \\
 4-460: asymmetric shape (both HI and optical image) \\
 4-461: HVC: one knot in one of the filaments that extends to the east of HI022823.7+244309 \\
 4-464: HVC: elongated filament \\
 4-473: parameters uncertain because of low S/N \\
 4-474: opt.id ambiguous; other possible counterpart AGC122812 (022606.5+255855) \\
 4-476: possible alternative opt.id with AGC122133 (022646.8+244236), but more than 2 arcmin from HI centroid \\
 4-483: HVC: bright central knot of system of filaments that extend throughout the grid \\
 4-486: opt.id ambiguous; other possible counterpart (or blend of emission) at 022908.9+251808. Parameters uncertain. \\
 4-487: HVC: one knot in one of the filaments that extends to the north of HI022823.7+244309 \\
 4-491: parameters affected by high noise; width determination uncertain \\
 4-499: alternative opt.id twice the distance away at 023528.2+253343 \\
 4-503: outer isophotes asymmetrical, distribution appears to stretch to the east. \\
 4-504: possible alternative opt. id with 024124.5+241627, similar galaxy but more than 1 arcmin away \\
 4-506: possible alternative opt. id with 024154.6+245657, but about 1 arcmin farther away from HI center \\
 4-523: HVC: small cloud \\
 4-528: parameters uncertain; signal partly blended with AGC120695 (025804.1+252656). Width could be larger that what is measured here. \\
 4-529: parameters uncertain; signal partly blended with AGC120678 (025759.1+252526) \\
 4-534: HI disk is elongated; appears affected by two nearby neighbors AGC120727 (025818.1+252657) and UGC2442. \\
 4-540: HI emission links this galaxy with UGC2457, ~10 arcmin north \\

\normalsize

\section{Objects of Note}

The catalog contains two objects worthy of some special attention. We report one non-HVC detection that cannot be confidently matched with an optical counterpart, as well as a peculiar High Velocity Cloud candidate detected at $+84$ \kms. Footnotes were made for each of these objects.

\begin{figure}[ht!]
\epsscale{0.9}
\plotone{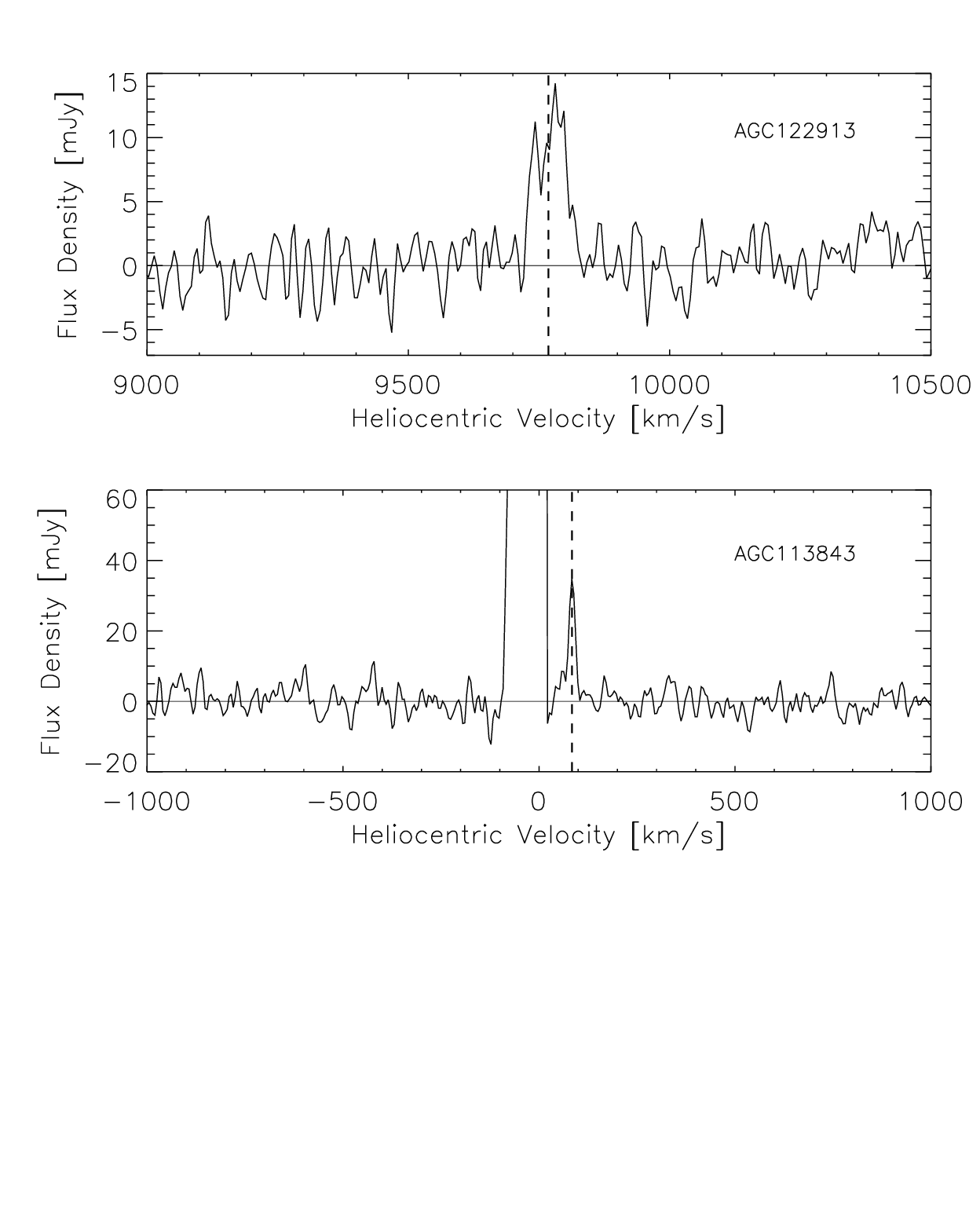}
\caption{Top panel: The spectrum for AGC122913 (object 449 in the catalog, HI021617.2+252616). The dashed line represents the heliocentric velocity measured for this source, 9678 \kms. The source, which is not aligned with any discernible optical emission in publicly available images, has a velocity width of 73 \kms and a S/N of 10.8, so we are very confident that the detection is real. Followup work may better determine its nature. Bottom panel: The spectrum for AGC113843 (object 362 in the catalog, HI011032.7+250559). The dashed line represents the heliocentric velocity measured for this source, 84 \kms. This object is identified as an HVC but its velocity is not consistent with other HVCs in the region. The source is extremely compact and has a narrow velocity width of 17 \kms, placing it on the border of ALFALFA's ability to detect it. Substantial polarization squint suggests that the source may be spurious, and followup observations will be necessary.\label{spectra}}
\end{figure}

The first of these, catalog object number 449 (HI021617.2+252616; AGC 122913), has no discernible optical emission in publicly available images. Its spectrum is shown in the top panel of Figure \ref{spectra}; the dashed line represents the measured heliocentric velocity. Point-source HI emission is detected at 9768 \kms \ with a width of 73 \kms. The object's profile is very well defined, and its S/N is 10.8. There is a good match between the two polarizations recorded in ALFALFA, and we are very confident that the detection is real. Nevertheless, there are no known galaxies within 2\arcmin, no identifiable optical emission to the depth of the DSS2 Blue images (this source is outside the SDSS footprint), and no continuum source or IRAS source counterpart, although the estimated B-band extinction in this area is $\sim 0.2$ magnitudes \citep{schlegel,burstein}. This source is a prime candidate for follow-up. Assuming that it is, indeed, an extragalactic HI source at this redshift, we estimate object 449 to be at a distance of 136 Mpc with an HI mass $\log (M_{HI}) = 9.56 M_{\odot}$. Given the velocity width of the profile and this HI mass, we would expect the optical images to show a face-on spiral system. Despite the lack of counterpart, there is no compelling evidence of the 1665.401/1667.358 MHz doublet of an OH megamaser. ALFALFA expects to detect several dozen OH megamasers \citep{alfalfa1}, which may explain some of the detections without optical counterparts; followup on candidate objects will be conducted by J. Darling (U. of Colorado).

Object number 362 in the catalog (HI011032.7+250559; AGC113843) is identified as an HVC, but it is found at a velocity of $+84$ \kms while the HVCs in this region are more typically found at heliocentric velocities near -400 \kms. Its spectrum is displayed in the bottom panel of Figure \ref{spectra}, again with the dashed line representing the measured heliocentric velocity of the source. The strong source near 0 \kms is the HI emission of the Galaxy. Detected with a SNR of 11.2, and with no potential optical counterpart visible in DSS2 Blue or SDSS DR7, this source's identification as an HVC is further evidenced by its spatial extent of $\sim 10$\arcmin, though its narrow velocity width of 17 \kms \ is on the border of ALFALFA's detection limit. There is significant polarization squint, so we caution that this signal may be spurious and followup will be necessary to determine the true nature of this source.

\section{Statistical Properties of the Catalog}

We first compare our catalog to previous HI and optical measurements. A portion of this region was covered in the Northern HIPASS catalog \citep{northhipass}, which added an extension in the range $+2^{\circ}<\delta<+25^{\circ}30^{\prime}$. HIPASS finds only 6 sources in this partial region of overlap. A more fair comparison can be found in \citet{alfalfa3}, since the northern HIPASS extension fully surveyed the area of the Virgo cluster. While ALFALFA found a total of 730 HI detections in that region, HIPASS reports only 40. Of the \ngal \ detections in Table \ref{params} that correspond to extragalactic objects, 296 ($65\%$) are new HI detections and 266 ($58\%$) are altogether new redshifts. This result, which is similar to that found by \citet{alfalfa3} and \citet{alfalfa6} in the highly-targeted Virgo Cluster region, demonstrates the failure of the ``conventional wisdom" of targeted HI surveys. Blind surveys are key to uncovering the true distribution of HI in the local Universe, and the sensitivity improvements of this second-generation survey further reveals the distribution of HI-rich galaxies.

Of the \ntot \ objects here, 90 (16$\%$) are High Velocity Clouds, 430 (80$\%$) are code 1 extragalactic detections, and 21 (4$\%$) are code 2 extragalactic detections. The median distance of extragalactic sources in the catalog is $\sim$114 Mpc, with a median heliocentric recessional velocity of $\sim$7300 \kms\ for all sources and $\sim$8300 \kms\ for all extragalactic sources. The High Velocity Clouds in this region possess highly negative velocities, ranging from -460 \kms \ to a single, unusual positive-velocity cloud candidate found at 84 \kms; -350 \kms\ is the median. The velocity width of clouds measured here is typically 25 \kms. Beyond the High Velocity Clouds, for which no optical match is made, there is one source in Table \ref{params} with no optical counterpart listed.

Relative to the overall ALFALFA sample, this catalog contains a large number of high velocity clouds, totaling 16$\%$ of detections. This includes some large filaments of high velocity gas, as well as compact and isolated knots. Paper V \citep{alfalfa5}, which surveys a region adjacent to this one, cointains a similar proportion, with 49 of 488 objects (10$\%$) identified as high velocity clouds of as bright knots of emission within a larger cloud complex. The anti-Virgo region was previously known to contain many HVCs, such as Wright's Cloud \citep{wright}. \citet{braun} completed a study of gas in this area, and found high velocity emission covering a total of 29$\%$ of an 1800 deg$^2$ region centered on $(\alpha,\delta) = (10^{\circ},35^{\circ})$ (J2000 coordinates). They identified $\sim$100 peaks of HVC emission with $cz_{\odot} \sim$0 to -350 km/s. Their map of the regions explored here agree with our finding that there is a great deal of gas and structure at high velocities. By contrast, in the Virgo and Leo regions \citep{alfalfa3,alfalfa6,alfalfa8}, the proportion of HVCs is smaller than 5$\%$ with a total of only $\sim$40 HVCs. A more thorough examination of the HVCs detected by ALFALFA and their distribution across the sky and in velocity space will be presented in a future paper. In particular, large and bright cloud complexes may require special data reduction, and compact, low-velocity width clouds may require a different source detection scheme.

\begin{figure}[ht!]
\epsscale{0.9}
\plotone{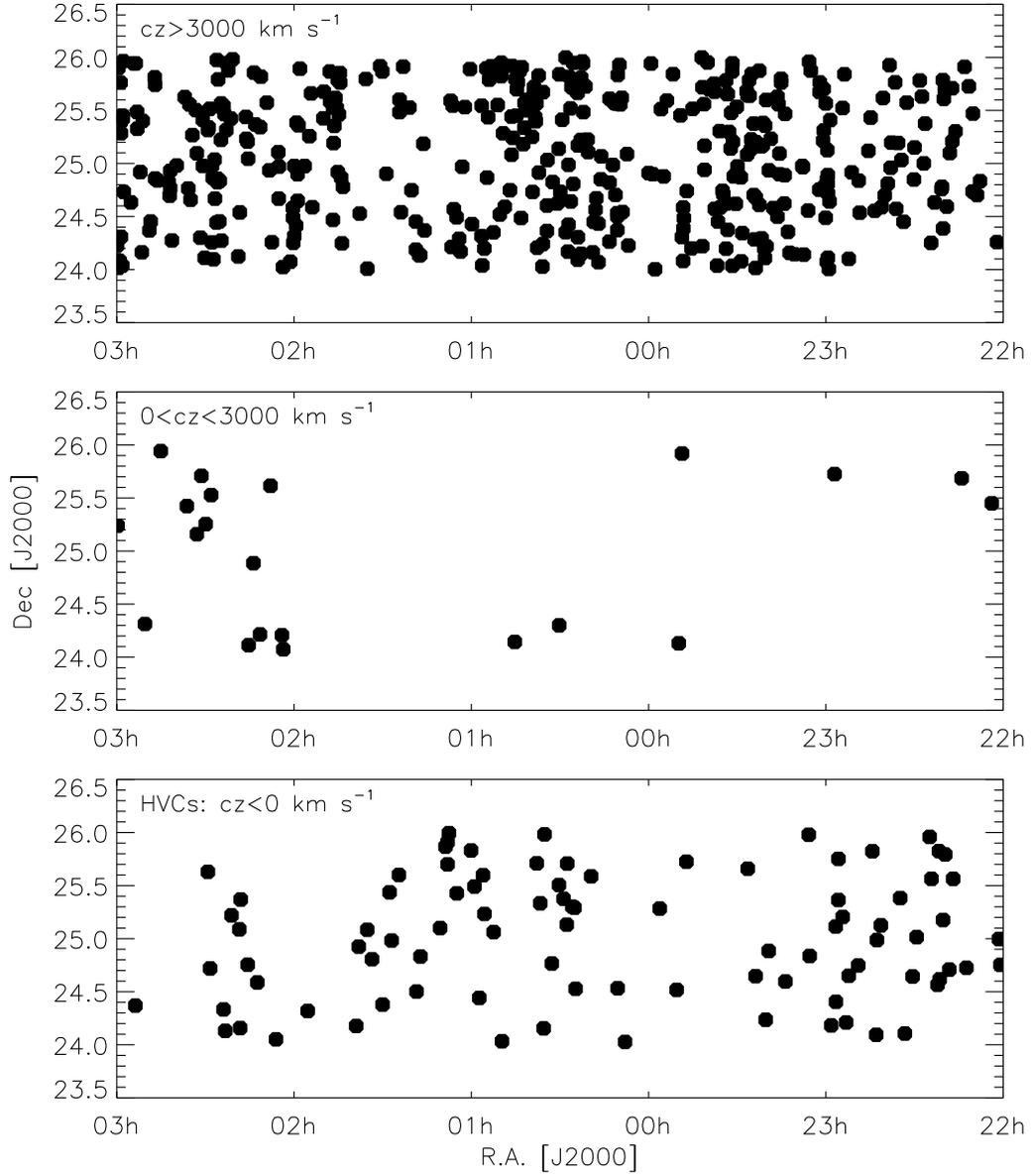}
\caption{Distribution of all the sources on the sky, in three different recessional velocity bins. The areas $\delta<24^{\circ}$ and $\delta>26^{\circ}$ are outside the catalog limits. The bottom panel represents the High Velocity Clouds.  \label{sky}}
\end{figure}

In Figure \ref{sky}, we show the sky distribution of the catalog, in three different velocity bins, revealing hints of large scale structure. The foreground underdensity discussed in \citet{alfalfa5} can be seen in the central panel, which includes sources with $0<cz_{\odot}<3000$ \kms, although there are some sources which appear at the low-declination end of the surveyed region. The source detection rate for this catalog is 3.3 sources per square degree. This is lower than the rate reported for previous ALFALFA catalogs in the spring sky \citep{alfalfa3,alfalfa6,alfalfa8} since these papers cover the Virgo cluster and Leo region, and thus we expect an overdensity in those catalogs. The bottom panel of Fig.\ref{sky} shows the distribution of the High Velocity Clouds.

\begin{figure}[ht!]
\epsscale{.85}
\plotone{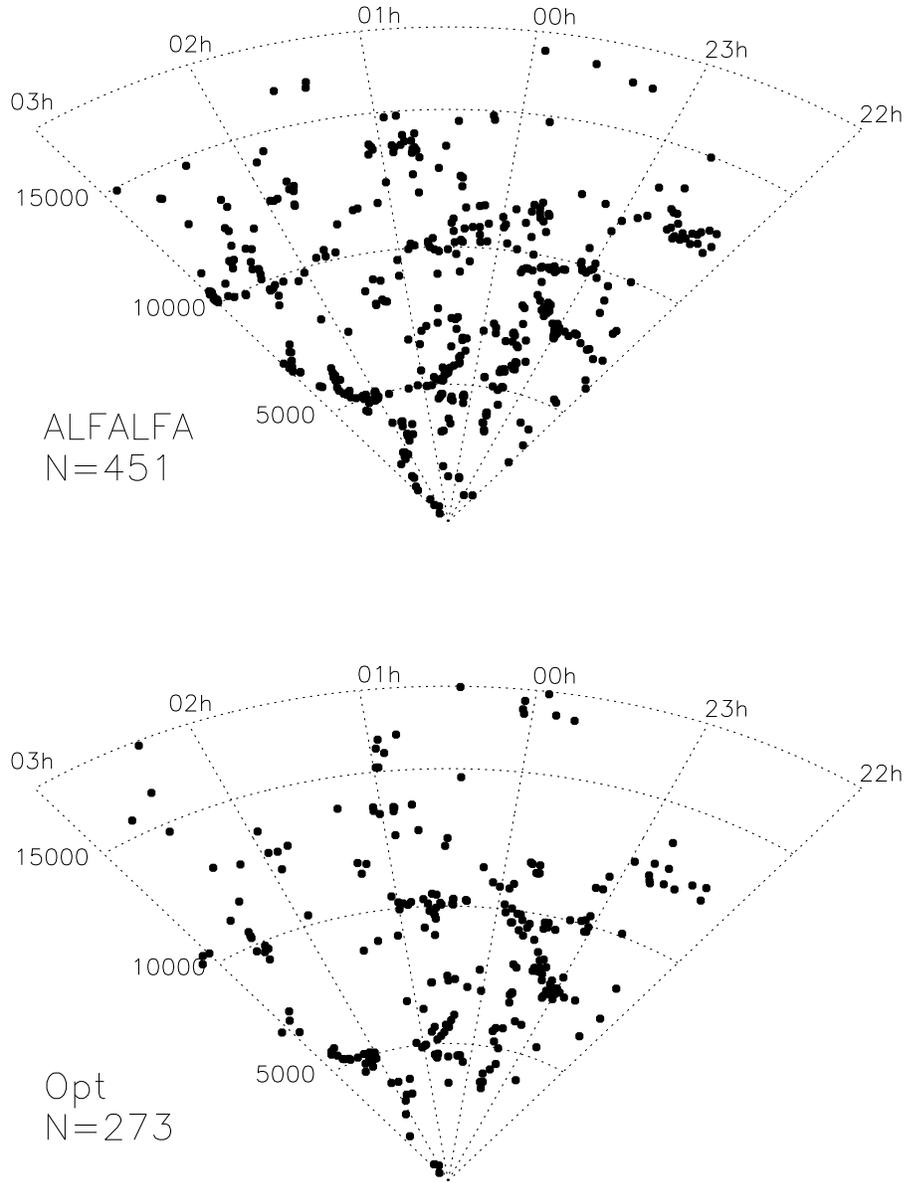}
\caption{Distribution of all the sources in the $2^{\circ}$ declination strip. The top panel shows all the ALFALFA detections presented here, while the bottom panel represents the galaxies with measured optical redshifts in the same volume of space. In each case, the number of plotted galaxies is reported. Note that due to RFI, ALFALFA is blind to cosmic emission between about 15000 and 16000 \kms. \label{cone}}
\end{figure}

In Figure \ref{cone} we present the redshift distribution of the extragalactic sources as a cone diagram. In the top panel, we show the distribution for this catalog; the bottom panel displays the previously known optical redshifts for the same region of sky. ALFALFA's inability to observe sources near 15,000 \kms\ is evident in the top panel of this diagram. Note that the decrement of sources at high velocity is due both to this broad source of RFI around 15,000 \kms\ that makes ALFALFA blind to cosmic emission there, as well as to decreased sensitivity to sources with distance. While there is a clear relationship between the structure probed through previous work in the optical and the ALFALFA detections, it is also clear that ALFALFA's detections in this region provide a substantial new dataset of redshift information. 

Figs. \ref{sky} and \ref{cone} also provide insight to the prominent local void in the foreground of the Pisces-Perseus Supercluster (PPS). \citet{alfalfa5} report no detections in ALFALFA or in previous optical surveys between $cz \sim 1000$ and $cz \sim 2500$ \kms\ in a 2$^{\circ}$ region centered on $\delta=+27^{\circ}$ in the range from $22^{h}00^{m}$ to $02^{h}00^{m}$. In the analogous region here, we do find several detections in ALFALFA, although the previous optical detections in the bottom panel of Fig. \ref{cone} agree with the findings of \citet{alfalfa5}. The void in the foreground of the PPS was reported by \citet{hg86} (see their Fig. 2), and while it is most prominent at higher declinations ($+30^{\circ}<\delta<+50^{\circ}$), the underdensity also stretches into $+20^{\circ}<\delta<+30^{\circ}$ and the regions reported in this paper and in \citet{alfalfa5}. 

\begin{figure}[ht!]
\epsscale{.6}
\plotone{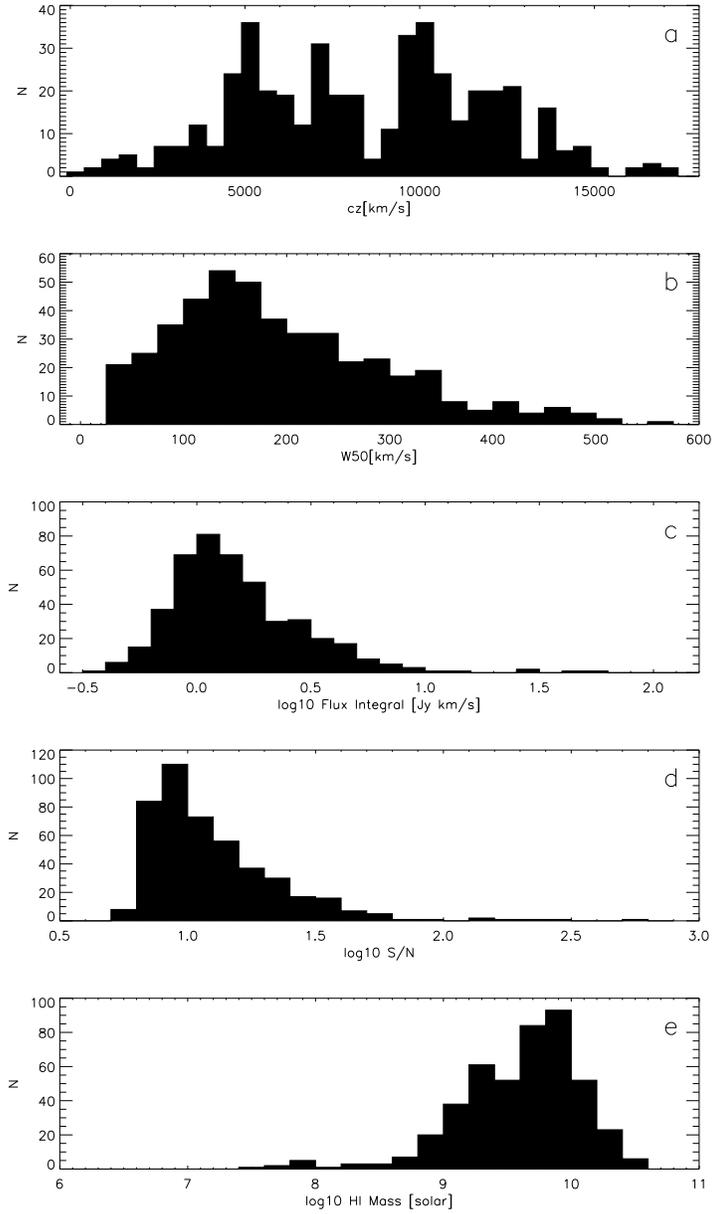}
\caption{Histograms of the HI detections with code 1 and 2 (i.e. excluding High Velocity Clouds): (a) heliocentric recession velocity in \kms; (b) HI line width at half power ($W50$) in \kms; (c) logarithm of the flux integral in Jy \kms; (d) logarithm of the signal-to-noise ratio; and (e) logarithm of the HI mass in solar units. \label{hist}}
\end{figure}

Figure \ref{hist} shows histograms displaying the main statistical properties of the catalog, including extragalactic detections of type 1 and 2. From top to bottom, the histograms represent the heliocentric velocity, velocity width, flux integral, $S/N$, and HI mass distributions. Fig. \ref{hist}a reiterates the few signals found near 15,000 \kms, as well as the large number of sources found in this catalog beyond the HIPASS velocity limit of 12,700 \kms \ \citep{hipass}. As shown in the last panel, Fig. \ref{hist}e, there are 8 sources with $M_{HI}<10^{8}M_{\odot}$, but all have masses $M_{HI}>10^{7.5}M_{\odot}$. This result is expected, since we find few sources very nearby, where ALFALFA is most sensitive to the lowest mass HI clouds.

\begin{figure}[ht!]
\epsscale{.6}
\plotone{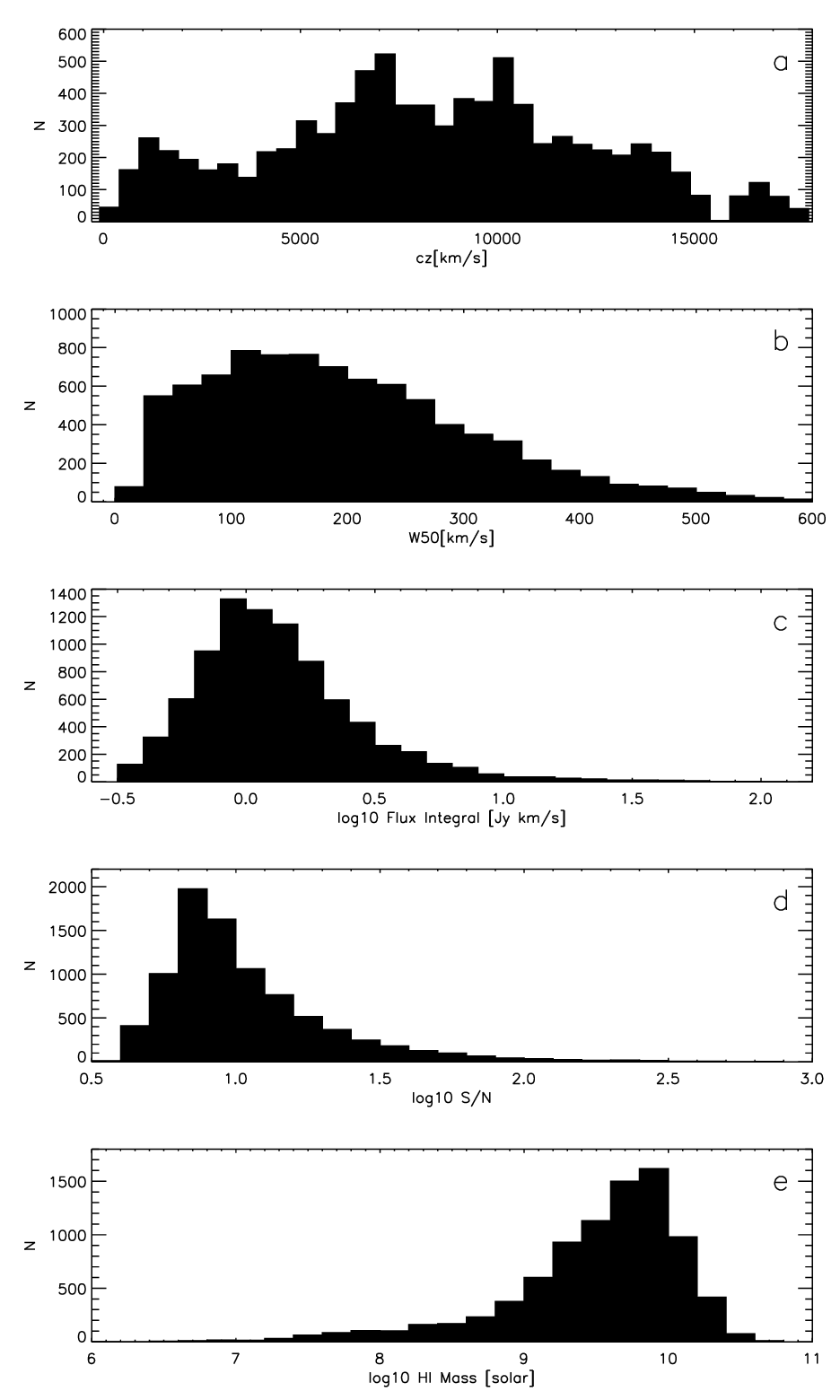}
\caption{Histograms of the total ALFALFA sample to date, including this work as well as previous publications in the Virgo region, Leo region, and anti-Virgo region at $\delta=+27^{\circ}$ \citep[]{alfalfa3,alfalfa5,alfalfa6,alfalfa8} and reduced data to be published in future data releases. This total sample includes all ALFALFA sources detected within the region of the sky having $22{\rm h}<\alpha<03{\rm h}$ and $24^{\circ}<\delta<28^{\circ}$ in addition to $07{\rm h}30{\rm m}<\alpha<16{\rm h}30{\rm m}$ and $4^{\circ}<\delta<16^{\circ}$. Plot includes HI detections with code 1 and 2 (i.e. excluding High Velocity Clouds): (a) heliocentric recession velocity in \kms; (b) HI line width at half power ($W50$) in \kms; (c) logarithm of the flux integral in Jy \kms; (d) logarithm of the signal-to-noise ratio; and (e) logarithm of the HI mass in solar units. \label{wholesamplehist}}
\end{figure}

Including this work, published ALFALFA catalogs to date contain 2,706 extragalactic HI detections over 653 deg$^2$, or 9$\%$ of ALFALFA's total survey area. Other ALFALFA catalogs include the Virgo Cluster region \citep{alfalfa3,alfalfa6}, the anti-Virgo region at a declination of 27$^{\circ}$ \citep{alfalfa5}, and the Leo region \citep{alfalfa8}. A series of histograms representing the total ALFALFA survey thus far are shown in Figure \ref{wholesamplehist}. This `20$\%$ ALFALFA Survey' sample includes this work as well as previous publications in the Virgo region, Leo region, and anti-Virgo region at $\delta=+27^{\circ}$ \citep[]{alfalfa3,alfalfa5,alfalfa6,alfalfa8} and reduced data to be published in future data releases. Figure \ref{wholesamplehist} thus represents all ALFALFA sources detected within the region of the sky having $22{\rm h}<\alpha<03{\rm h}$ and $24^{\circ}<\delta<28^{\circ}$ in addition to $07{\rm h}30{\rm m}<\alpha<16{\rm h}30{\rm m}$ and $4^{\circ}<\delta<16^{\circ}$. This combined catalog of published and soon-to-be published sources includes 7,166 extragalactic objects coded as type 1, and 1,462 coded as type 2. Soon, the total publicly available ALFALFA catalog will surpass HIPASS in source counts. 

Generally speaking, the sample presented in this work is consistent with the ALFALFA catalog as a whole, as summarized by Figure \ref{wholesamplehist}. The velocity distribution in Fig. \ref{wholesamplehist}a has smoothed out much of the large scale structure, but in particular the Virgo Cluster at $ cz_{\odot}\le 3000$ \kms \ remains an obvious feature. Once again, the lower density of sources at large velocities is caused by a combination of our reduced sensitivity as a function of distance, as well as sources of rfi such as the FAA radar, which essentially leaves us blind to emission at 15,000 \kms. ALFALFA's sensitivity to sources at the Virgo Cluster's distance, where we can detect HI masses down to 10$^{7}$\msun, also extends the HI mass distribution (Fig. \ref{wholesamplehist}e) to lower masses than the complementary histogram for this catalog alone (Fig. \ref{hist}e). Overall, the 20$\%$ ALFALFA Survey includes $\sim$300 sources with neutral hydrogen masses below 10$^8$\msun. The paucity of sources at low heliocentric velocity (and therefore small distances) in the +25$^{\circ}$ catalog reduces the number of the lowest-mass sources detected, but as a whole ALFALFA has clearly demonstrated an ability to probe the distribution below 10$^{8}$\msun in the local Universe. 

A comparison of Fig. \ref{hist}d with Fig. \ref{wholesamplehist}d shows that the overall ALFALFA sample includes more sources with lower S/N than the catalog in this work. This can be explained by the difference in objects coded as 1 and those coded as 2, since objects with code 2 may have a lower signal-to-noise ratio as long as they have a corroborating redshift from optical data. In the Virgo and Leo regions, which have been well-studied and are covered by the Sloan Digital Sky Survey, there are therefore more objects with lower S/N and code 2, while the catalogs in the fall sky (this work and \citet{alfalfa5}) are mainly composed of objects coded as type 1.

With this catalog and its predecessors, ALFALFA continues to demonstrate its utility and success as a second-generation blind HI survey. With coverage now totaling roughly 9$\%$ of the final survey area and including $\sim$2700 extragalactic HI detections and many High Velocity Clouds, ALFALFA will soon outpace the first-generation efforts of HIPASS. These results will contribute to our understanding of the distribution of neutral hydrogen in the local Universe, the mass function, and the dependence on environment of galaxy parameters. For the first time, a cosmologically significant sample of HI sources, unbiased with respect to the stellar populations of host galaxies, will soon be available.

\acknowledgements
This work was supported by NSF grants AST-0607007 and AST-9397661, and by grants from the National Defense Science and Engineering Graduate (NDSEG) fellowship and from the Brinson Foundation. We acknowledge the use of NASA's {\it SkyView} facility (http://skyview.gsfc.nasa.gov) located at NASA Goddard Space Flight Center, as well as the Sloan Digital Sky Survey Data Release 7 (http://www.sdss.org). Funding for the SDSS and SDSS-II has been provided by the Alfred P. Sloan Foundation, the Participating Institutions, the National Science Foundation, the U.S. Department of Energy, the National Aeronautics and Space Administration, the Japanese Monbukagakusho, the Max Planck Society, and the Higher Education Funding Council for England. This research has also made use of the NASA/IPAC Extragalactic Database (NED) which is operated by the Jet Propulsion Laboratory, California Institute of Technology, under contract with the National Aeronautics and Space Administration.

\bibliographystyle{apj}
\bibliography{myreferences}

\clearpage

\clearpage

\clearpage

\clearpage

\clearpage

\end{document}